\documentclass[a4paper,11pt]{article} 
\usepackage{amsmath}
\usepackage{epsfig,float}
\usepackage{color}
\usepackage{subfigure}
\usepackage{float}
\usepackage{enumitem}
\usepackage[]{hyperref}
\usepackage[normalem]{ulem}
\usepackage{setspace} 

\usepackage{graphicx}
\usepackage{amsmath, amssymb, amsthm}
\usepackage{latexsym}
\usepackage{color}
\usepackage{hyperref}
\usepackage{graphicx}
\usepackage{subfigure}
\usepackage{float}
\usepackage{csquotes}
\usepackage{lineno}

\allowdisplaybreaks

\usepackage{amssymb}
\usepackage[ruled, ]{algorithm2e}
\usepackage{booktabs}
\usepackage{xurl}

\begin{document}

\centerline{\LARGE \bf
Effect of Vaccine Dose Intervals: Considering Immunity }

\medskip

\centerline{\LARGE \bf Levels, Vaccine Efficacy, and Strain Variants for}

\medskip

\centerline{\LARGE \bf Disease Control Strategy}

\color{black}
\medskip

\vspace*{1cm}

\centerline{\bf Samiran Ghosh$^1$, Malay Banerjee$^{1}$, Amit K Chattopadhyay$^{2,*}$}

\vspace{0.5cm}

\centerline{ $^1$ Indian Institute of Technology Kanpur, Kanpur - 208016, India}

\centerline{$^2$
Department of Applied Mathematics and Data Science, Aston University,}
\centerline{Aston Centre for Artificial Intelligence Research and Applications (ACAIRA),}
\centerline{Aston Triangle, Birmingham B4 7ET, UK}

\vspace{0.5cm}
\centerline{$^*$ Corresponding author: Amit K Chattopadhyay}

\vspace{2cm}

\noindent
{\bf Abstract.}
In this study, we present an immuno-epidemic model to understand mitigation options during an epidemic break. The model incorporates comorbidity and multiple-vaccine doses through a system of coupled integro-differential equations to analyze the epidemic rate and intensity from a knowledge of the basic reproduction number and time-distributed rate functions. Our modeling results show that the interval between vaccine doses is a key control parameter that can be tuned to significantly influence disease spread. We show that multiple doses induce a hysteresis effect in immunity levels that offers a better mitigation alternative compared to frequent vaccination which is less cost-effective while being more intrusive. Optimal dosing intervals, emphasizing the cost-effectiveness of each vaccination effort, and determined by various factors such as the level of immunity and efficacy of vaccines against different strains, appear to be crucial in disease management. The model is sufficiently generic that can be extended to accommodate specific disease forms.

\vspace{1cm}


\medskip

\vspace*{0.5cm}

\setcounter{equation}{0}
\setcounter{section}{0}
\setcounter{page}{1}

\section{Introduction}
In recent decades, the global population has faced a series of viral outbreaks, each leaving a significant impact on health, society, and the economy. These include the SARS epidemic of 2002-2003 \cite{SARS_2002_1,xing2010anatomy}, the H5N1 influenza outbreak in 2005 \cite{H5N1_1,H5N1_2}, the H1N1 influenza pandemic in 2009 \cite{H1N1_1,smith2009origins}, the Ebola crisis in 2014 \cite{ebola_1,holmes2016evolution}, and currently, the ongoing and unpredictable COVID-19 pandemic \cite{chattopadhyay2021infection,cachon2020systematic}, which has persisted for more than two years. These successive epidemics have posed substantial challenges to public health, social structures, and economic systems worldwide.

The field of epidemic modeling has witnessed significant advancements, expanding beyond simple models to embrace more intricate frameworks. Noteworthy developments include the evolution towards multi-compartment models \cite{brauer_compartmental,martcheva,tolles2020modeling}, models with time-varying or nonlinear disease transmission rates \cite{hethcote_nonlinear_incidence_1,chavez_nonlinear_incidence_1}, multipatch models \cite{multipatch_1,multipatch_2,ghosh2023understanding}, agent based models \cite{kerr2021covasim,hoertel2020stochastic}, and multigroup models \cite{kuniya_1,li2010global}. Furthermore, the exploration of spatiotemporal aspects has led to the formulation and study of spatiotemporal models, as evidenced by \cite{chang2022sparse,banerjee2023spatio}. \textcolor{black}{Some other analyzes on the dynamics of SIR models can be found in the literature \cite{zhao2018stability,zhang2023dynamical}.} For a comprehensive exploration of these topics, one can refer to monographs such as \cite{martcheva,brauer,capasso2008mathematical}.

Vaccination stands as a cornerstone in combatting the spread of infections, with established vaccines for diseases like measles, typhoid, poliomyelitis, tuberculosis, malaria, Haemophilus influenzae type-b (Hib), Japanese encephalitis, pneumococcal disease, meningococcal meningitis, and the more recently developed COVID-19, according to the World Health Organization \cite{pipeline_vaccine_WHO}. Concurrently, ongoing efforts are directed towards developing vaccines and monoclonal antibodies for diseases such as enterotoxigenic Escherichia coli, herpes simplex virus, shigella, norovirus, and improved influenza \cite{pipeline_vaccine_WHO}. As vaccination programs progress, critical questions arise regarding the availability of vaccines, the rate at which vaccine efficacy diminishes, and the optimal dosage regimen, especially in scenarios of limited vaccine supply. The focus shifts to determining the optimal interval between consecutive vaccine doses, a crucial consideration to avoid unnecessary doses and the associated hysteresis effect, thereby optimizing the use of vaccines and financial resources. {\color{black}{In the context of the Oxford/AstraZeneca COVID-19 (AZD1222) vaccine groups, it has already been shown that accurate implementation of dosage intervals plays a key role in vaccine efficacy \cite{yang_liu_2022}. This is somewhat complementary to optimizing vaccine dosing schedules that can prolong immunity and suppress the emergence of new variants of SARS-CoV-2 potentially due to intermittent lapses in protection \cite{dogra_2023}. None of these approaches though incorporate the crucial comorbidity factor that is the central premise of this study, in conjunction with dosage interval moderation.}}

Determining a suitable and optimal gap between two consecutive vaccinations plays a pivotal role in controlling an epidemic and minimizing economic costs for several reasons. First and foremost, the timing of vaccine doses directly influences the development and sustainability of immunity within the population. A well-calibrated gap allows for the optimal buildup of immune responses, providing a more robust shield against the spread of infectious diseases. Moreover, finding the right interval between vaccine doses is crucial for preventing the resurgence of infections. Administering doses too closely may lead to a diminishing effect, potentially causing a hysteresis effect, where the immunity gained from the first dose interferes with the effectiveness of the subsequent doses. On the other hand, spacing doses too far apart may leave individuals vulnerable to infection during the gap, reducing the overall efficacy of the vaccination strategy. {\color{black}{Another less discussed issue is the {\em in vitro} versus {\em in vivo} modes of operation: a vaccine developed under laboratory conditions may not be as effective in real-life situation. More importantly, with other variants, e.g. the Delta variant, the booster dose may lose potency faster than anticipated, facts that have been recently modeled \cite{menegale_2023}.}}

From an economic perspective, the cost associated with vaccination programs is a significant consideration. Frequent vaccination incurs not only direct costs related to the purchase and administration of vaccines but also indirect costs tied to the disruption of regular activities, strain on healthcare resources, and potential economic losses due to illness and workforce absenteeism. Therefore, optimizing the vaccination gap is crucial in striking a balance between widespread infection and the financial investment required for effective immunization programs.

Compartmental epidemic models, where each compartment is determined by the daily number of cases and disease transmission rates, recovery, and death rates are distributed over the time-since-infection, offer a more accurate depiction of epidemic progression compared to classical SIR-type models \cite{bmb_our,jmb,SG_MB_VV_MMNP}. In this study, we present an epidemic model where all parameters are distributed over time following an infection, thereby accounting for the dynamically changing level of immunity in the population. The level of immunity in the population is influenced by the acquired immunity of the recovered and multiple vaccine doses administered in the population at specified intervals. Additionally, we introduce a comorbid compartment with higher infectivity, implicitly capturing the impact of multiple strains on determining the number and height of epidemic peaks within a specific time interval. We estimate all relevant time-distributed parameters using available clinical and experimental data. Our modeling results reveal that the coexistence of multiple strains can influence the frequency and height of epidemic outbreaks. Furthermore, our findings indicate that frequent vaccine administration may not be necessary and could be substituted by an effective immunity memory build-up through a hysteresis effect. The optimal gap between two consecutive vaccines should be determined by analyzing the level of immunity in the population. To address this, we formulate an optimal control problem aimed at minimizing both the number of infections and the economic costs related to vaccination. The primary objective of this work is to achieve a clear understanding of the optimal gap between two consecutive vaccinations, a dimension that has not been thoroughly explored in the existing literature.

The article is structured as follows: In Section \ref{section2}, we formulate the model incorporating time-distributed parameters and the dynamic level of immunity, additionally introducing a model that considers the impact of multiple strains. The calculation of the basic reproduction number is presented in the same section. Section \ref{section3} is dedicated to the estimation of relevant time-distributed rate functions, using available clinical and experimental data. The results and findings, including the influence of multiple strains and the observed hysteresis effect resulting from the gap between consecutive vaccine doses, are discussed in Section \ref{section4}. 

\section{Model formulation}\label{section2}

\subsection{\bf Basic Model}
We consider a population with four compartments: susceptible individuals ($S(t)$), infected individuals ($I(t)$), recovered individuals ($R(t)$) and dead individuals ($D(t)$). It is assumed that the sum of these compartments remains constant and equals the total population $N$, i.e., 
\begin{equation}\label{n0}
 S(t)+I(t)+R(t)+D(t)=N,\;\;\text{for all} \;\; t \geq0.
\end{equation}
Let $J(t)$ denote the number of newly infected individuals at time $t$, which is governed by the following equation
\begin{equation}\label{n1}
  \frac{dS(t)}{dt} = - J(t). 
\end{equation}
The classical SIR-type epidemic models assume that the number of newly infected individuals is proportional to the product of the number of infected $I(t)$ and the number of susceptible $S(t)$ at time $t$. However, in more realistic scenarios, this assumption is not true, and the disease transmission rate depends upon time-since-infection for the infected individuals \cite{bmb_our}. Since the viral load dynamics within an infected individual varies with time-since-infection, the infectivity of the infected individuals also varies with time-since-infection \cite{SG_MB_VV_MMNP}. Here, we consider the number of new infections determined by time-since-infection dependent transmission rate, given by the following equation:
\begin{equation}\label{n2}
   J(t) =\frac{S(t)}{N} \int_0^t \beta(t-\eta) J(\eta) d \eta. 
\end{equation}
We assume here that the infection transmission rate at time $t$ from the individuals $J(\eta)$ infected at time $\eta$ depends on the time difference $t-\eta$. In the case of respiratory viral infections, it depends on the viral load in the upper respiratory tract. We will specify the function $\beta(t)$ below. 

Similarly, the number of newly recovered $R_n(t)$ and dead individuals $D_n(t)$ can be described by the following equations:
\begin{equation}\label{n3}
R_n(t) \equiv \frac{d R(t)}{dt} = \int_0^t r(t-\eta) J(\eta) d \eta ,\,\, \;\; D_n(t) \equiv \frac{d D(t)}{dt} = \int_0^t d(t-\eta) J(\eta) d \eta .
\end{equation}
Distributed recovery and death rates $r(\eta)$ and $d(\eta)$ assume the probability of recovery and death as functions of time-since-infection $\eta$. They are determined from the immunological (clinical) data (see Section 3). 

Finally, differentiating equality (\ref{n0}) and taking into account (\ref{n1}) - (\ref{n3}), we obtain the  equations for $S(t)$ and $I(t)$ as follows: 
\begin{equation}
\label{n5}
  \frac{d S(t)}{dt} = - \frac{S(t)}{N} \; \int_0^t \beta(t-\eta) J(\eta) d\eta ,
\end{equation}
\begin{equation}
\frac{d I(t)}{dt} =  \frac{S(t)}{N} \; \int_0^t \beta(t-\eta) J(\eta) d\eta -   \int_0^t r(t-\eta) J(\eta) d \eta - \int_0^t d(t-\eta) J(\eta) d \eta . \label{n6}
\end{equation}
Completing them by equations (\ref{n1}) and (\ref{n3}), we obtain the formulation of an immuno-epidemiological model with distributed infectivity, recovery and death rates.


\subsection{\bf Vaccination and immunity}\label{level_immunity}
The level of immunity in the population plays an important role in combating the progression of the epidemic. This level of immunity can vary over time based on several important factors, such as the rate of vaccination, time-post-vaccination-dependent vaccine effectiveness, and time-post-recovery-dependent acquired immunity.

To account for the influence of vaccination on epidemic progression, we introduce a new variable $m(t)$ corresponding to the immunity level in the population. First, we analyze the relationship of the immunity level with vaccination. This is hypothesized through the constitutive relation
\begin{equation}
m(t) = \frac{1}{N} \int_0^t \phi(t-\eta) V'(\eta) d\eta ,
\end{equation}
where $V(t)$ is the number of vaccinated individuals at time $t$, and $V'(t)$ is the rate of vaccination,  the function $\phi(t)$ describes how immunity changes with time. It is a positive function with $\phi(0)=1$ (if a vaccine is initially fully efficient), otherwise $\phi(0)>0$, then it increases up to some maximal value and decreases after that due to immunity waning. {\color{black}{Overall, we assume that $0<\phi(t)<1$.}}

\subsubsection{Multiple Vaccination Doses}
Now, we incorporate the impact of multiple vaccine doses in the population at various intervals. The expression for the immunity level $m(t)$ is then provided as follows:
\begin{equation}\label{m2}
  m(t) = \frac{1}{N} \left( \sum_{i=1}^K \int_0^t \phi_i(t-\eta) V_i'(\eta) d\eta  \right).
\end{equation}
Here $V_i(t)$ and $\phi_i(t)$ denote the number of vaccination and the efficacy of vaccines respectively corresponding to all the doses, starting from $i=1$. $K$ is the total number of doses administrated in the population. Suppose that two consecutive vaccine doses $i$ and $i+1$ are administrated with a time gap $T_{i,i+1}$, and the first dose was started from $t=T_0$. Then for any $i \in \{1,2, \cdots, K\}$, we have
\begin{equation}\label{s20}
  V_i(t)= \left\{
  \begin{array}{cc}
  0,& t < T_0+ T_{1,2}+T_{2,3}+ \cdots + T_{i-1,i}\\
  &\\
  \text{positive},& t \geq T_0+ T_{1,2}+T_{2,3}+ \cdots + T_{i-1,i}
  \end{array} \right..
\end{equation}
 Although the effectiveness of the initial vaccine dose and subsequent booster doses may differ, for the sake of simplicity, we assume uniform efficacy for all vaccine doses in this study. Our interest in this part of the analysis is to understand whether multiple booster doses compound the effects of the previous rounds of vaccination or reach a plateau beyond which they are ineffective.\\

\subsubsection{Acquired immunity of recovered}
Next, consider the acquired immunity of the recovered individuals. 
Then the effective immunity at time $t$ coming from both the vaccination of healthy susceptible $S$ and the infection acquired immunity is given by:

\begin{eqnarray}\label{m3}
  m_1(t) &=& \frac{\alpha}{N} \left( \sum_{i=1}^K \int_0^t \phi_i(t-\eta) V_i'(\eta) d\eta  \right) + \frac{(1-b)}{N}{\int_0^t \psi(t-\eta) R_n(\eta) d\eta}, \nonumber\\
\end{eqnarray}
and the effective immunity at time $t$ coming from vaccination of comorbid class $P(t)$ is given by:

\begin{eqnarray}\label{mm3}
  m_2(t) &=& \frac{1-\alpha}{N} \left( \sum_{i=1}^K \int_0^t \phi_i(t-\eta) V_i'(\eta) d\eta  \right) +  \frac{b}{N}{\int_0^t \epsilon \psi(t-\eta) R_n(\eta) d\eta},\nonumber\\
\end{eqnarray}
where, $0< \epsilon <1 $ is a constant, $\alpha$ is the proportion of healthy susceptible $S$ who are newly vaccinated. \textcolor{black}{The value of $\epsilon<1$ signifies that the infection acquired immunity for the comorbid individuals is less than that for the non-comorbid individuals.} Parameter $b$ characterizes the proportion of comorbidity among the newly recovered individuals while the function $\psi(t)$ describes how acquired immunity changes over time.
We choose functions $\phi_j$ and $\psi_j$ focusing on multiple criteria, prioritizing the acquisition-fading function, exponential fading function and the power law function. The functional fits are parameterized against epidemiological data comprising the epidemic form. Assuming homogeneity, we consider $\phi_j \equiv \phi$ and $\psi_j \equiv \psi$, for $j=1, 2, \cdots, K$. Moreover, to account for \enquote{bounded} infection growth, we impose a constraint on the population immunity level: $0<m_1(t)+m_2(t)<1$.

\subsubsection{Impact of Multiple Strains}
Immunity in the population corresponds to the decrease of the number of susceptible individuals. As such, instead of equality (\ref{n0}), we have
\begin{equation}\label{m4}
 S(t)   = N - (I(t) + D(t) + P(t)+ m_1(t) N).
\end{equation}
Note that $m_2(t)$ does not appear in the equation \eqref{m4} because $m_2(t)$ is the level of immunity in the comorbid class; it is only considered later in the equation for $P(t)$, where $P(t)$ is the predisposed comorbid population. Let, $J_1(t)$ represents the number of daily new infection in the healthy susceptible class $S(t)$ and $J_2(t)$ represents the number of daily new cases in the comorbid class $P(t)$. Then the total number of daily new cases is given by $$J(t)=J_1(t)+J_2(t),$$ where,
$$J_1(t)=\frac{S(t)}{N} \int_0^t \beta(t-\eta) J(\eta) d \eta,$$
$$J_2(t)=\frac{\kappa P(t)}{N} \int_0^t \beta(t-\eta) J(\eta) d \eta,$$
$\kappa>1$ is a constant, \textcolor{black}{which signifies that the comorbid individuals are more prone to the infection as compared to the non-comorbid individuals}.

\begin{figure}[ht!]
\begin{center}
	\mbox{\includegraphics[width=11cm,height=5cm]{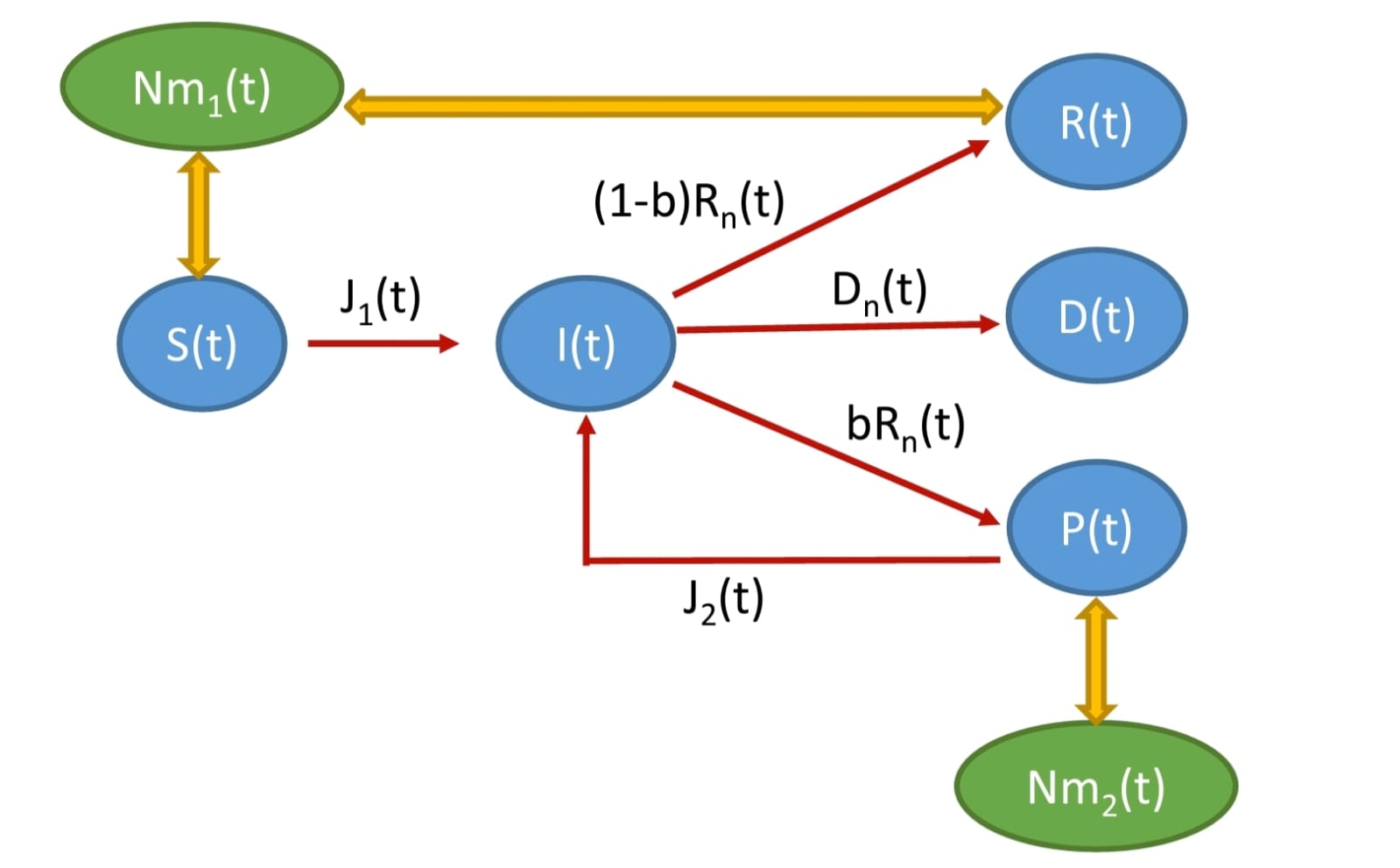}}
\caption{Schematic diagram for the system (\ref{m3})-(\ref{m6}).}
\label{fig_diag_1}
\end{center}
\end{figure}

The governing equation for the comorbid class $P(t)$ is given by
\begin{equation}
P(t) =  b \int_0^t R_n(\eta) d \eta - \int_0^t J_2(\eta)d \eta - m_2(t) N.\label{n2}
\end{equation}
The equation for the infected compartment is given by
\begin{equation}
  \frac{d I}{dt} =  \textcolor{black}{\frac{S(t)+\kappa P(t)}{N} \int_0^t \beta(t-\eta) J(\eta) d \eta} -   \int_0^t r(t-\eta) J(\eta) d \eta - \int_0^t d(t-\eta) J(\eta) d \eta . \label{m5}
\end{equation}
 Finally the equations for the recovered and death compartments are 
\begin{equation}
   \frac{d R}{dt}   =  (1-b)R_n(t), \;\;\;
\frac{d D}{dt}  = \int_0^t d(t-\eta) J(\eta) d \eta \;(= D_n(t)), \label{m6}
\end{equation}

where, $$R_n(t)=\int_0^t r(t-\eta) J(\eta) d \eta,$$ which represents the daily sum of completely recovered individuals (with proportion $(1-b)$) and comorbid individuals (with proportion $b$). We obtain complete model (\ref{m3})-(\ref{m6}) with distributed infection, recovery and death rates, and population immunity. The corresponding flowchart is shown in Figure~\ref{fig_diag_1}.

\subsection{\bf Basic reproduction number}

In the beginning of epidemic, assume that $I=P=D=m_1=0$. Then, using (\ref{n1}), we can write equation (\ref{n5}) in the following form:
\begin{equation}
  S'(t) =  \frac{S}{N} \; \int_0^t \beta(t-\eta) S'(\eta) d\eta . \label{s5}
\end{equation}
Suppose, $S(t) = N - \epsilon e^{\lambda t}$. Then from the above equation we get,
\begin{eqnarray}
\frac{dS(t)}{dt} &=& -\epsilon \lambda e^{\lambda t} = \frac{N-\epsilon e^{\lambda t}}{N}\int_0^t \beta(t-\eta) (-\epsilon \lambda) e^{\lambda \eta} d\eta.
\end{eqnarray}
Now equating the terms with the first power of $\epsilon$ in both sides we get:
\begin{eqnarray}
\label{s6}
e^{\lambda t} &=& \int_0^t \beta(t-\eta) e^{\lambda \eta} d\eta \nonumber \\
\Rightarrow\;\;\;\; 1 &=& \int_0^t \beta(t-\eta)e^{-\lambda(t-\eta)} d\eta.
\end{eqnarray}
The dispersion relation (the relation involving the parameters that determine the stability) can be obtained by setting $\lambda=0$ and the dispersion relation is given by:

$$ \int_0^t \beta(t-\eta)  d\eta = 1,$$
which can be written as
$$   \int_0^t \beta(x)  d x = 1. $$
If we assume that $\beta(x) > 0$ for $0 \leq x \leq \tau$ and $\beta(x) = 0$ for $ x > \tau$, \textcolor{black}{where, $\tau$ is assumed to be the average disease duration}. Then we can define the basic reproduction number as
\begin{equation}\label{r0}
 \mathcal{R}_0 = \int_0^\tau \beta(x)  d x. 
\end{equation}
Then $\lambda$ in (\ref{s6}) is positive (epidemic growth) if and only if $\mathcal{R}_0>1$.

\paragraph{\bf Note} The derivation of basic reproduction number remains unaltered even if we assume a reproduction substituting of the form $S(t) = N-\epsilon a^{\lambda t}$, where $a>0$.


\section{Parameter Estimation}\label{section3}

\subsection{\bf Statistical toolbox}
The estimations and curve fittings are done by minimizing the Sum of Squared Errors (SSE). For the curve fitting to data we mainly use the `Curve Fitting Toolbox', which is a collection of graphical user interfaces (GUIs) and M-file functions built on the MATLAB technical computing environment \cite{curve_fitting_toolbox}.
The toolbox provides the fitted curve along with the goodness of fit. Gamma distributions are estimated using the inbuilt function {\it{fitdist(:,gamma)}} in MATLAB. This function is used to fit a vector of data $X=(x_1, x_2, \cdots , x_n)$ by a gamma distribution of the form $\frac{1}{b^a \Gamma (a)}x^{a-1}e^{-x/b}$, where $a$ and $b$ are the shape and scale parameters. This function gives the maximum likelihood estimators of $a$ and $b$ for the gamma distribution which are the solutions of the simultaneous equations 
$$\log \hat{a} -\Psi(\hat{a})= \log \bigg( \bar{X}/ \big( \prod_{i=1}^{n} x_i \big)^{1/n} \bigg),$$
$$\hat{b} = \bar{X}/ \hat{a},$$
where $\bar{X}$ is the sample mean of the data $X$ and $\Psi$ is the digamma function given by
$$\Psi(x)=\Gamma'(x)/\Gamma(x).$$
The function {\it{fitdist(:,gamma)}}  estimates the shape and scale parameters with $95\%$ confidence interval.\\

\subsection{\bf Choice of vaccination function $V(t)$}
We assume that the vaccination function $V(t)$ that is started at time $t=t_0$, follows the function 
\begin{equation}\label{formula_vaccination}
  V(t)= \left\{
  \begin{array}{cc}
  0,& t < t_0\\
  L \left[ N-(N-V_0) e^{-k(t-t_0)}\right],& t \geq t_0
  \end{array} \right..
\end{equation}
where, $N$ is the total population size, $L$ is the proportion of the population expected to be vaccinated, $k$ is the rate of vaccination, $V_0$ is the number of vaccination at time $t=t_0$, and the associated parameter values are listed in Table.~\ref{table:estimated}.

\begin{figure}[ht!]
\begin{center}
		\mbox{\subfigure[]{\includegraphics[scale=0.28]{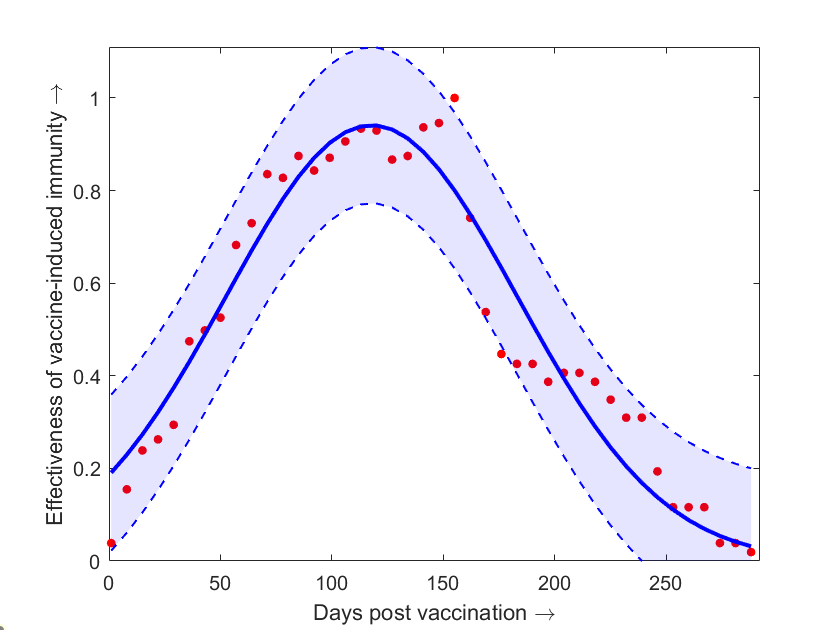}}
    \subfigure[]{\includegraphics[scale=0.28]{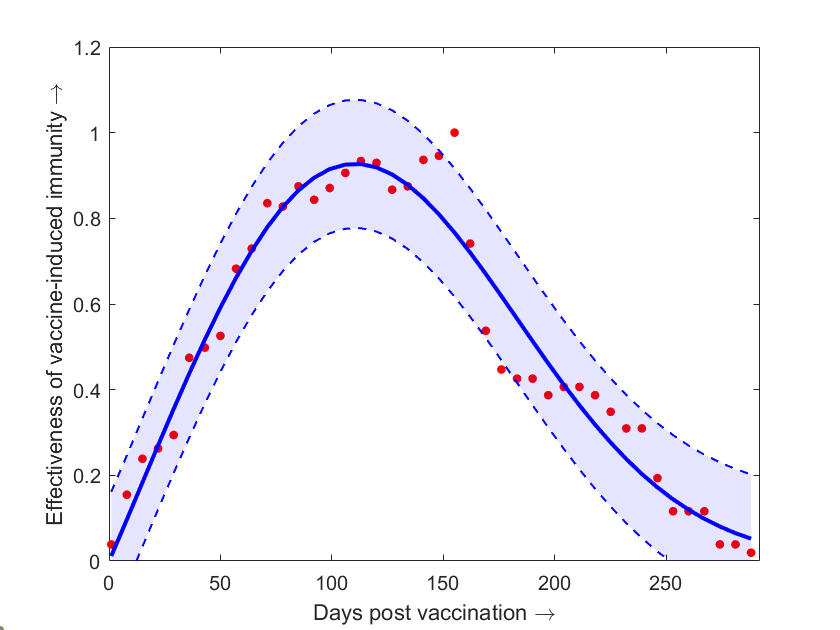}}}
   
\caption{ The effectiveness of vaccine-induced immunity  $\phi(t)$ as a function of the days post-vaccination (data source- \cite{SG_MB_VV_MMNP,vaccine_effectiveness}). \textcolor{black}{The shaded regions represent the $95\%$ confidence interval of the fitted function in the epidemiologically feasible region.} (a) corresponds to the formula \ref{formula_phi}; (b) corresponds to the formula \ref{formula_phi1}.}
\label{fig1}
\end{center}
\end{figure}

\subsection{\bf Estimation of $\phi(t)$ and $\psi(t)$}
Due to the lack of availability of {\color{black}{sector data, that is  separate data incorporating the impacts of comorbidity and the ones without,}} we take an initial simplifying step by assuming that $\phi_j \equiv \phi$ and $\psi_j \equiv \psi$, for $j=1, 2, \cdots, N$. We use the data of vaccine-induced immunity from \cite{SG_MB_VV_MMNP} and fit (least square fitting) the function $\phi(t)$ as follows (see Figure~\ref{fig1}a,b)

\begin{equation}\label{formula_phi}
\phi(t)=a_1 e^{-\left(\frac{t-b_1}{c_1}\right)^2},
\end{equation}
where, $a_1=0.9411$ with $95\%$ CI $(0.8886,0.9937)$, $b_1=117.8$ with $95\%$ CI $(113.5, 122)$, and $c_1=92.44$ with $95\%$ CI $(86.06, 98.82)$ (Figure~\ref{fig1}a). The goodness of fit is as follows:
SSE=$0.2807$, R-square=$0.9308$, Adjusted R-square=$0.9273$ and RMSE=$0.08483$.\\
Now we fit the same data of vaccine-induced immunity with a stretched power-law function given by

\begin{equation}\label{formula_phi1}
\phi(t)=a_2 t^{b_2} e^{-c_2 t^{d_2}},
\end{equation}
where, $a_2=0.01152$ with $95\%$ CI $(-.001766, 0.02481)$, $b_2=1.023$ with $95\%$ CI $(0.719, 1.328)$, $c_2=5.01 \times 10^{-6}$ with $95\%$ CI $(-1.398 \times 10^{-5}, 2.4 \times 10^{-5})$, $d_2=2.412$ with $95\%$ CI $(1.747, 3.077)$ (Figure~\ref{fig1}b). The goodness of fit is as follows:
SSE=$0.2218$, R-square=$0.9453$, Adjusted R-square=$0.941$ and RMSE=$0.0764$. {\color{black}{It must be noted that death due to natural causes versus death due to infection have two different timescales of operation; the former is way more protracted than the latter, an aspect that plays a major role in ascribing average values to death rates in comorbidity models.}}

We observe that the stretched power law function \eqref{formula_phi1} gives better fitting to the data as compared to the Gaussian function \eqref{formula_phi}. To check the robustness of the choice of the stretched power law function \eqref{formula_phi1} we compared the goodness of fit with other possible candidates such as the Gaussian function and, acquisition-fading function.

\begin{figure}[ht!]
\begin{center}
	   \includegraphics[scale=0.4]{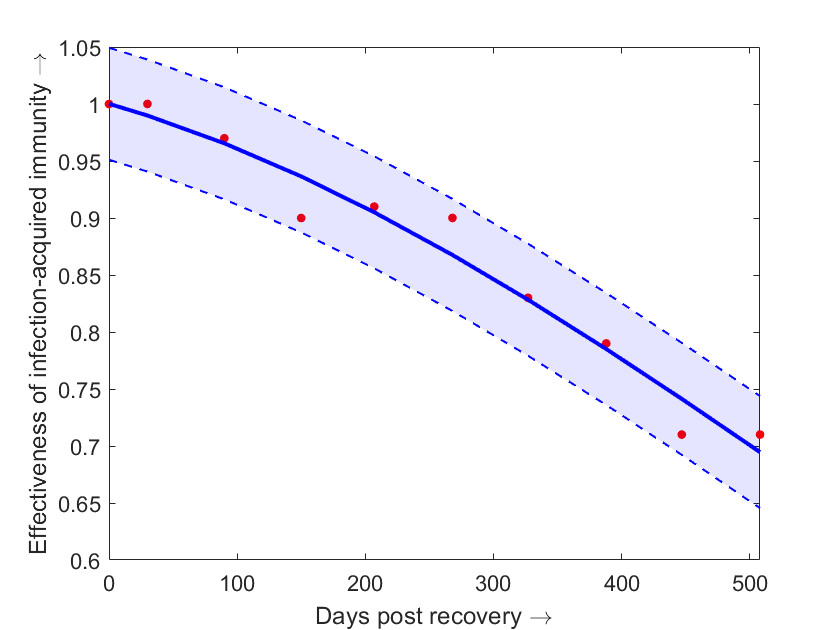}
\caption{The effectiveness of infection-acquired immunity  $\psi(t)$ as a function of the days post recovery (data source- \cite{recovered_effectiveness,SG_MB_VV_MMNP}).The blue dots are the real data and the red curves are the functions fitted to the data. \textcolor{black}{The shaded region represents the $95\%$ confidence interval of the fitted function.} The details of the fitted parameter values are given in the text.}
\label{fig2}
\end{center}
\end{figure}

The function for the effectiveness of acquired immunity $\psi(t)$ is fitted to the data available in \cite{recovered_effectiveness}, by the following function (Figure~\ref{fig2}):
\begin{equation}\label{formula_psi}
\psi(t)=a_3 e^{-\left(\frac{t-b_3}{c_3}\right)^2},
\end{equation}
where, $a_3=1.035$ with $95\%$ CI $(0.8742, 1.195)$, $b_3=-206.6$ with $95\%$ CI $(-704.3, 291)$ and $c_3=1133$ with $95\%$ CI $(500.3, 1765)$.


\begin{figure}[ht!]
\begin{center}
		\mbox{\subfigure[]{\includegraphics[scale=0.42]{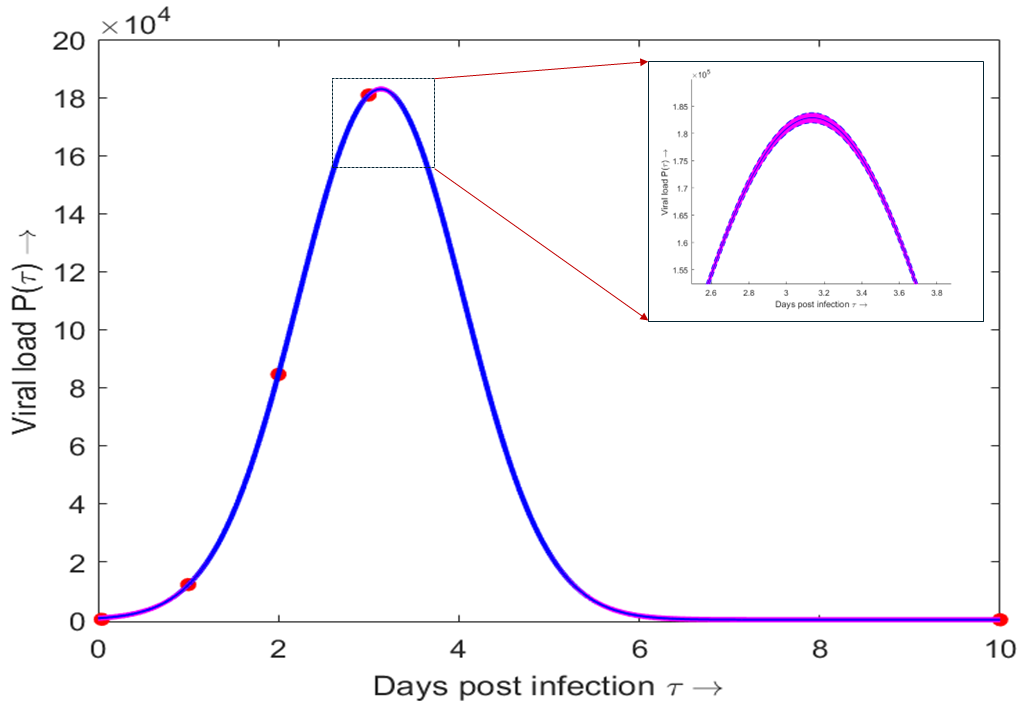}}
	}
\caption{Viral load as a function of the days post infection. The blue dots are the real experimental data for Omicron variant taken from \cite{chan2021sars}. The red curve is the gamma function fitted to the blue dots. \textcolor{black}{The shaded region represents the $95\%$ confidence interval of the fitted function.}  The details of the fitted parameter values are given in the text.}
\label{fig2viral}
\end{center}
\end{figure}

\subsection{\bf Estimation of transmission rate $\beta(\tau)$.}
We assume that the transmission rate $\beta(\tau)$ is proportional to the viral load $P(\tau)$, i.e., $\beta(\tau)=c P(\tau)$, where, $c$ is a proportionality constant and it depends upon the transmission rate (which mainly depends on the behavioral aspects and not on the virus variants) between infected and susceptible individuals. We fit the function $P(\tau)$ with the experimental data of viral load depending on the number of hours post-infection as available in \cite{chan2021sars}. In \cite{chan2021sars}, the authors experimented to understand the viral replication kinetics of SARS-CoV-2 variants in ex vivo cultures of the human respiratory tract and the experiment was performed up to $71$ hours-post-infection. Also, we assume that after $10$ days of the days-post-infection, the viral load becomes negligible \cite{quilty2022test}. Using all these information, we fit $P(\tau)$ by the following function (Figure~\ref{fig2viral}):
\begin{equation}\label{formula_viral_load}
P(\tau)=a_4 e^{-\left(\frac{\tau-b_4}{c_4}\right)^2},
\end{equation}
where, $a_4=1.829 \times 10^5$ with $95\%$ CI $(1.805 \times 10^5, 1.852 \times 10^5)$, $b_4=3.136$ with $95\%$ CI $(3.073, 3.2)$ and $c_4=1.294$ with $95\%$ CI $(1.234, 1.353)$. \\

\begin{table}[ht!]
\begin{center}
\caption{Parameter values}
\begin{tabular}{llll}
\hline\\[-0.8em]
{Parameters} & Description & Estimated value & Source \\
\hline\\[-0.8em]
$N$ & Total population & $10^7$ & -  \\
$V_0$ & Initial number of  & $500$ &- \\
 &  vaccination &  & \\
$L$ & proportion of population& $0.75$&-\\
 &  to be vaccinated & &\\
  &  expected & &\\
 &  to be vaccinated & &\\
$k$ & rate of vaccination & $0.002,0.003,$ &-\\
 &  & $0.005$ &\\
$c$ & proportionality constant &$0.44 \times 10^{-5}$ &-\\
$\alpha$ & proportion of vaccination & $0.8$ &-\\
 &  among susceptible &  &\\
$\kappa$ & proportionality constant & $1.1$ &-\\
$b$ & rate of comorbidity & $0.2$ & -\\
$\epsilon$ & -& $0.7$ & -\\
$\phi(t)$ & effectiveness of vaccine& Eqn.~\ref{formula_phi} & \cite{SG_MB_VV_MMNP,vaccine_effectiveness}\\
 & -induced immunity & (Figure~\ref{fig1}a) & \\
$\psi(t)$ & effectiveness of infection & Eqn.~\ref{formula_psi} & \cite{SG_MB_VV_MMNP,recovered_effectiveness}\\
 & -acquired immunity & (Figure~\ref{fig1}b) & \\
$P(\tau)$ & Viral load & Eqn.~\ref{formula_viral_load} & \cite{SG_MB_VV_MMNP,chan2021sars,quilty2022test}\\
&  &  (Figure~\ref{fig2}) & \\
$r(t), d(t)$ &  and death rates & Eqn.~\ref{formula_recovery_death}  & \cite{SG_MB_VV_MMNP}\\
 &  and death rates & (Figure~\ref{fig_recovery_death}) & \\
$V(t)$ & total vaccination function & Eqn.~\ref{formula_vaccination} & -\\
\hline
\end{tabular}
\label{table:estimated}
\end{center}
\end{table}

\subsection{\bf Estimation of $r(t)$ and $d(t)$}
\begin{figure}[ht!]
\begin{center}
		\mbox{\subfigure[]{\includegraphics[scale=0.4]{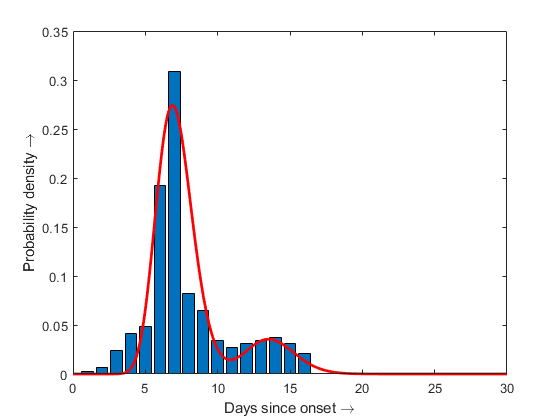}}
			\subfigure[]{\includegraphics[scale=0.4]{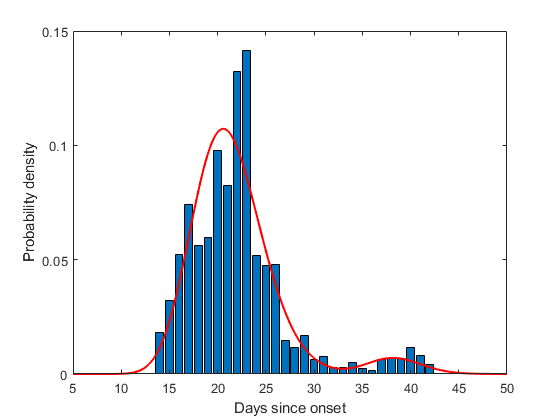}}}
\caption{Time-distributed rate functions of (a) recovery and (b) death as functions of days post the onset of infection. The red curves show the best fitted bimodal gamma distributions ( Ref.~\cite{SG_MB_VV_MMNP}).} 
\label{fig_recovery_death}
\end{center}
\end{figure}

In the literature on epidemic modelling, the choice of gamma distributions to model distributed recovery period is well known \cite{Bailey,chowell_book,Lloyd}. However, the use of bimodal gamma distributions in epidemic modeling can indeed provide a more accurate representation of the recovery or death rate functions when there are distinct groups with different time intervals. From a linear combination of two different gamma distributions, we can capture the variability in the recovery or death times more effectively. The recovery and death distributions used in \cite{SG_MB_VV_MMNP} and are given by:
\begin{equation}\label{formula_recovery_death}
r(t)=p_0\mathcal{F}_1(t),\;\;\;\; d(t)=(1-p_0) \mathcal{F}_2(t),
\end{equation}
where,
$$\mathcal{F}_1(t)=\frac{0.85}{b_1^{a_1} \Gamma (a_1)} t^{a_1-1} e^{-\frac{t}{b_1}} + \frac{0.15}{d_1^{c_1} \Gamma (c_1)} t^{c_1-1} e^{-\frac{t}{d_1}} $$
with estimated parameter values $a_1=32.17136$, $b_1=0.2206$, $c_1=65.40545$, $d_1=0.210$ (Figure~\ref{fig_recovery_death}a) and 
$$\mathcal{F}_2(t)=\frac{0.94}{b_2^{a_2} \Gamma (a_2)} t^{a_2-1} e^{-\frac{t}{b_2}} + \frac{0.06}{d_2^{c_2} \Gamma (c_2)} t^{c_2-1} e^{-\frac{t}{d_2}} $$
with estimated parameter values  $a_2=36.02855$, $b_2=0.57511$, $c_2=140.11379$, $d_2=0.27636$ (Figure~\ref{fig_recovery_death}b). $p_0$ is the survival probability which is assumed to be $p_0=0.9975$.
All other parameter values are listed in the Table.~\ref{table:estimated}.

\section{Results and findings}\label{section4}
\subsection{\bf Impact of multiple strains}
\begin{figure}[ht!]
\begin{center}
		\mbox{\subfigure[]{\includegraphics[scale=0.6]{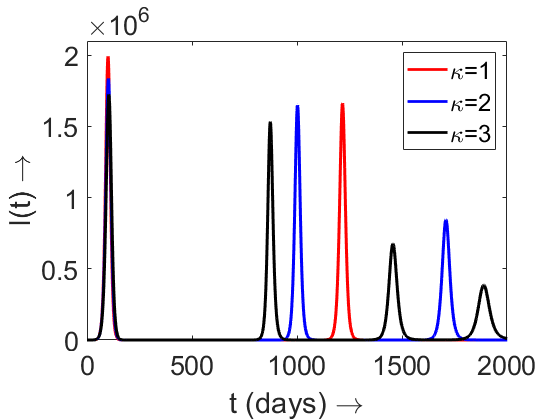}}
			}
\caption{Plot of $I(t)$ for different choice of $\kappa$. The associated parameter values are chosen as estimated before and as in Table.~\ref{table:estimated}.}
\label{multiple_strain_1}
\end{center}
\end{figure}

In this section, we study the influence of the existence of multiple strains on the epidemic progression in a population. Here the parameter $\kappa$ accounts for the existence of multiple strains. A higher value of $\kappa$ implies the co-existence of prominent strains with very different transmission rates. In Figure~\ref{multiple_strain_1}, we plot $I(t)$ for different choices of $\kappa$. We observe that as the value of $\kappa$ increases, the number of epidemic peaks also increases and the peaks appear relatively frequently. This finding points to multiple infection waves for epidemics driven by multiple strains compared to a single wave for single-strained infections. However, the maximum height of individual peaks is seen to decrease as has been recently observed with the Covid second and third waves \cite{chattopadhyay2021infection}. 

\begin{figure}[ht!]
\begin{center}
		\mbox{\subfigure[]{\includegraphics[scale=0.28]{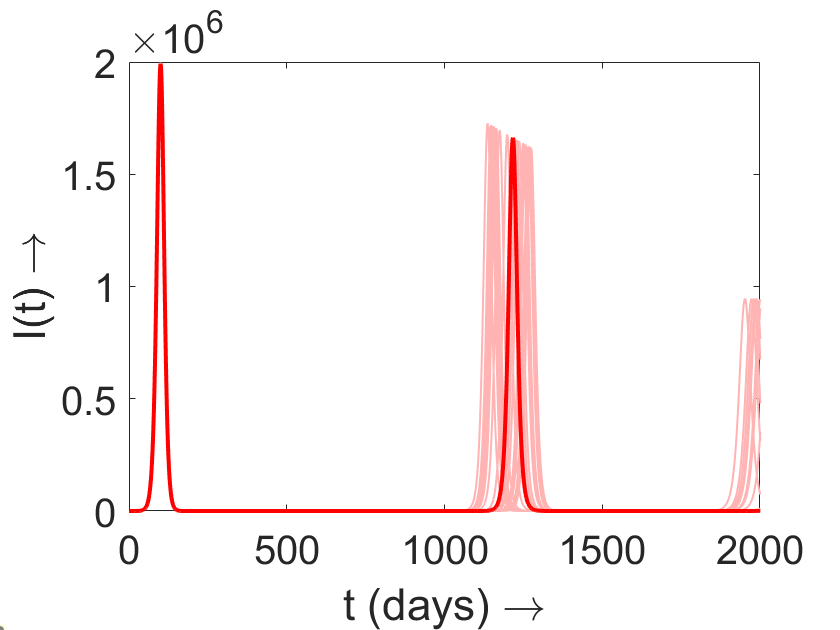}}
  \subfigure[]{\includegraphics[scale=0.28]{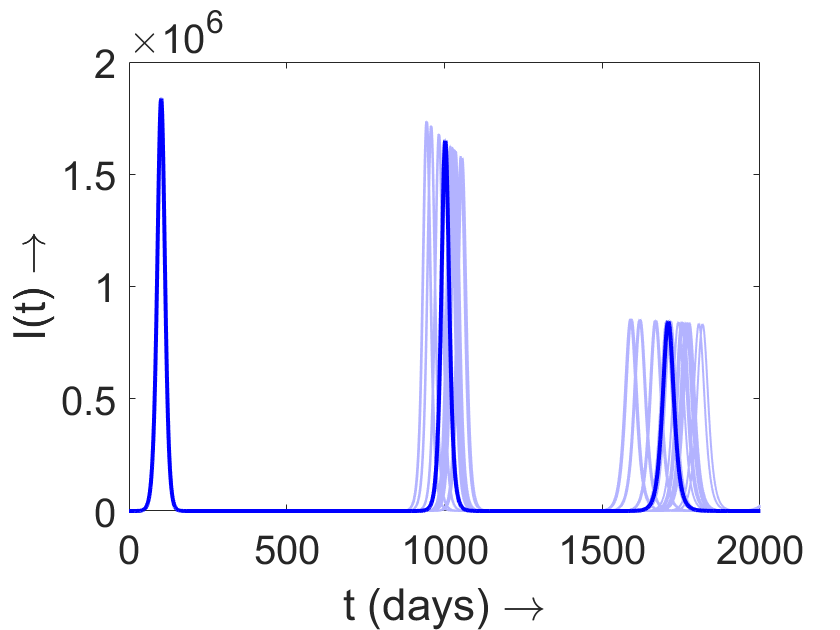}}}
  \mbox{
  \subfigure[]{\includegraphics[scale=0.28]{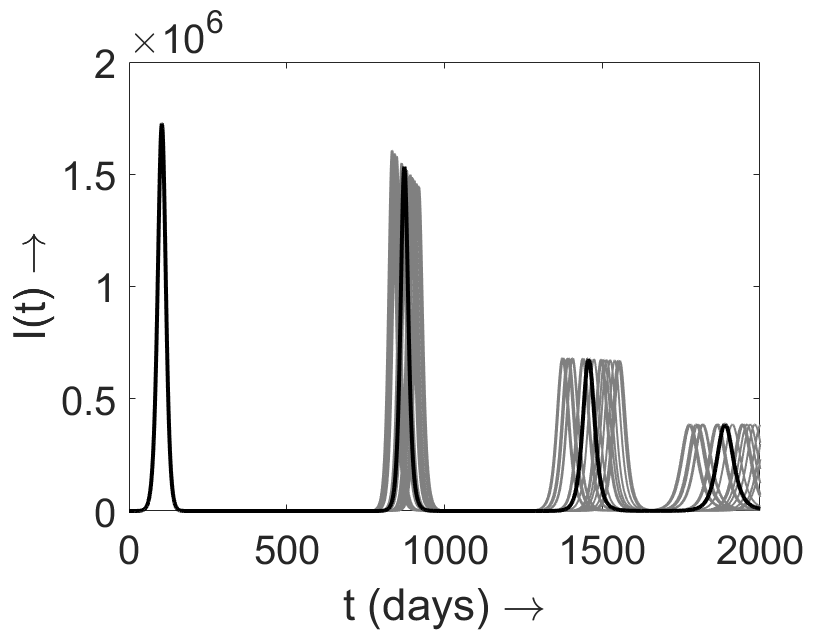}}
			}
\caption{Plot of $I(t)$ for different choice of $\kappa$. \textcolor{black}{(a) $\kappa=1$; (b) $\kappa=2$; (c) $\kappa=3$. The bold colored curves correspond to $c_3=1133$ in formula \eqref{formula_psi}. The light colored curves correspond to $20$ randomly chosen values of $c_3$ in the interval $[1033,\; 1233]$. } The other associated parameter values are chosen as estimated before and as in Table.~\ref{table:estimated}.}
\label{multiple_strain_1_sensitivity}
\end{center}
\end{figure}

\textcolor{black}{
To explain the sensitivity of the model parameters on the model outcome 
$I(t)$, we randomly chose 20 values of 
$c_3$ (in formula \eqref{formula_psi}) within the interval 
$[1033,1233]$. The outcome $I(t)$ is shown in Fig.~\ref{multiple_strain_1_sensitivity}. In this figure, the bold color curves correspond to $c_3=1133$, while the light color curves correspond to the 20 randomly chosen values of $c_3$. This simulation result demonstrates that the model outcome $I(t)$ is sensitive to the model parameters, although the principal trend of the outcome remains largely unchanged. Another important observation is that the first epidemic peak is not sensitive to the parameter $c_3$. This is because, at the beginning of the epidemic, the acquired immunity function (as described in \eqref{formula_psi}) is not very influential, and as the epidemic progresses, this acquired immunity function causes variations in $I(t)$. Similar sensitivity analyzes were performed for other parameters, but the main trend of the outcome remained nearly the same.}

\subsection{\bf Effect of interval between successive vaccine doses: Hysteresis Effect}

In this section, we investigate the effect of the gap between two consecutive doses of vaccination. We assume a time range  from $0$ to $T=2000$ days (which is almost 5.5 years), where we're trying to control an epidemic. To explain the effect we consider three scenarios as follows:
\paragraph{Scenario-1:} Vaccine doses administered with $4$ months gap.
\paragraph{Scenario-2:} Vaccine doses administered with $8$ months gap.
\paragraph{Scenario-3:} Vaccine doses administered with $12$ months gap. \\

\begin{figure}[ht!]
\begin{center}
		\mbox{\subfigure[]{\includegraphics[scale=0.4]{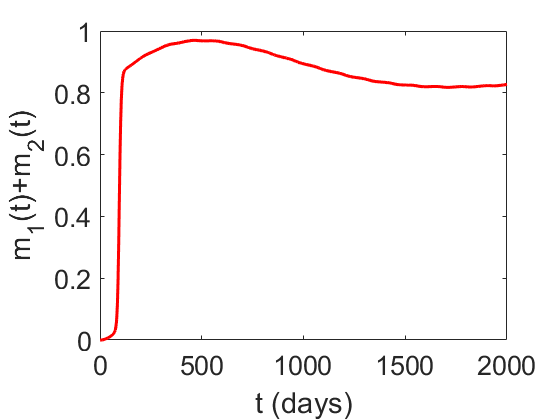}}
			\subfigure[]{\includegraphics[scale=0.4]{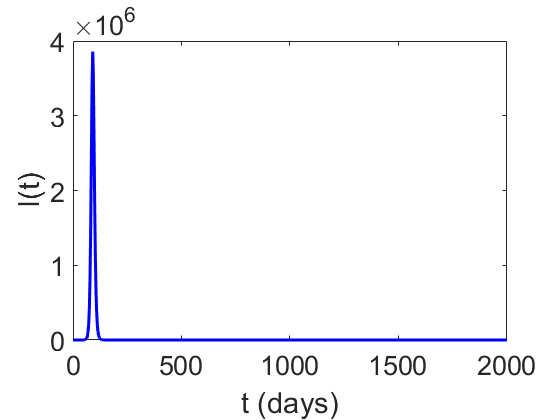}}}
   \mbox{\subfigure[]{\includegraphics[scale=0.4]{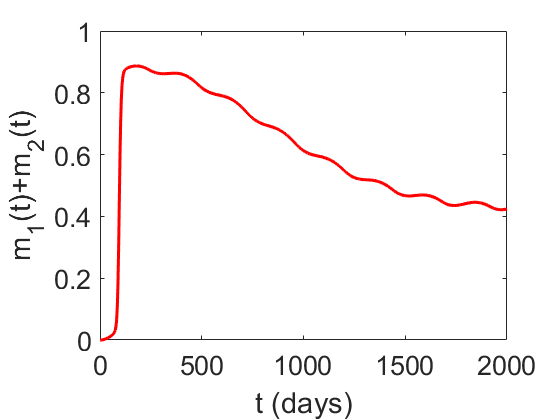}}
			\subfigure[]{\includegraphics[scale=0.4]{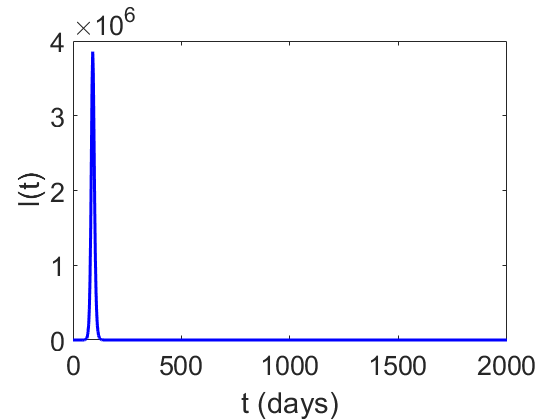}}}
    \mbox{\subfigure[]{\includegraphics[scale=0.4]{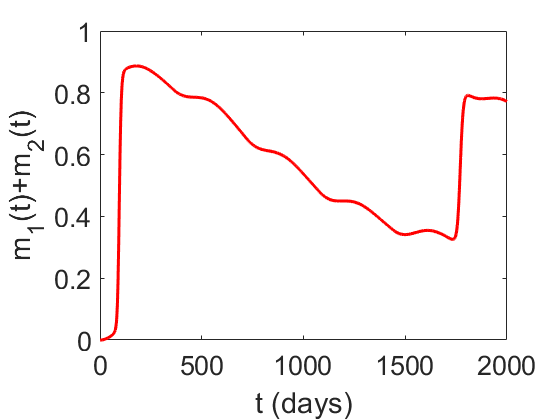}}
			\subfigure[]{\includegraphics[scale=0.4]{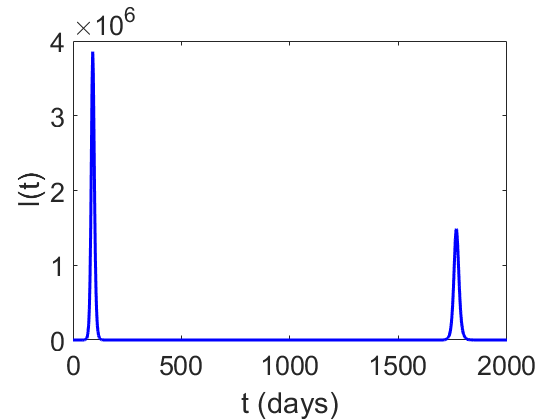}}}
 
\caption{(a), (b) correspond to Scenario-1; (c), (d) correspond to Scenario-2; (e), (f) correspond to Scenario-3. The associated parameter values are chosen as estimated before and as in Table.~\ref{table:estimated}.}
\label{gap_comparison}
\end{center}
\end{figure}

\noindent
For simplicity, we assumed that each vaccine dose has the same efficacy. From Figure~\ref{gap_comparison}, we observe that Scenario-1 and Scenario-2 depict almost the same epidemic progression whereas Scenario-3 depicts a different type of progression. The result shows that instead of \textcolor{black}{administrating the vaccine with a gap of $4$ months, a vaccination spanning a gap of $8$ months produces the same type of epidemic progression, though the level of immunity is slightly less}. Also, we note that Scenario-1 requires repeated vaccinations compared to Scenario-2 within the period $0$ to $2000$ days, but both scenarios eventually accord the same level of immunity. This is a key observation that can help us to avoid unnecessary vaccinations. On the other hand, from Figure~\ref{gap_comparison}(e), (f), we observe that if the vaccination gap is larger (i.e., $1$ year in this case), then consecutive epidemic peaks can appear in future. Thus a proper gap should be maintained to minimize future epidemic outbreaks. The summary of these observations is the need to exercise optimal control. 

\subsection{\bf Optimization problem}

\begin{figure}[ht!]
\begin{center}
		\includegraphics[scale=0.5]{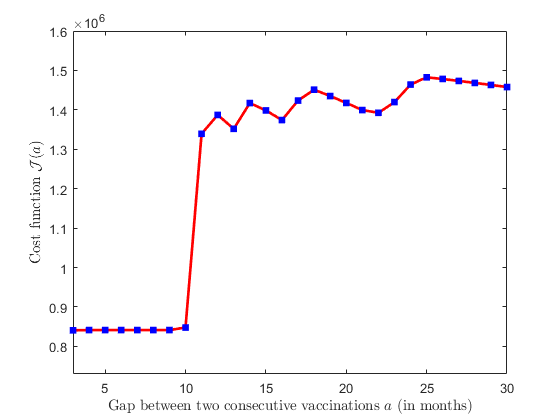}
			
\caption{Plot of cost function $\mathcal{J}(a)$ for $c=0.01$, $d=5$, and all other parameter values are chosen as estimated before and as in Table.~\ref{table:estimated}.}
\label{cost_function_1}
\end{center}
\end{figure}

Based on the previous numerical results, we can consider the following optimization problem:

\begin{equation}\label{cost_function_n}
\mathcal{J}(n) = \min_{n \in \mathbb{N},\;\; 0 \leq m(t;n) \leq1} \;\; c \int_0^T I(t;n) dt +d n,
\end{equation}
where $T>0$ is the maximum time we consider. $n$ is the number of vaccination campaigns administrated in the population during the time interval $[0,\;T].$ We assume that $T=an$, for some $a>0$, i.e., two vaccination campaigns are considered with a gap of $a$ time units. $d$ is a positive constant that accounts for the cumulative cost per vaccination campaign. \textcolor{black}{$m(t;n)$ and $I(t;n)$ denote the level of immunity and number of infected at time $t$ for a given $n$, respectively.} \textcolor{black}{$I(t;n)$} is the solution of our model for a particular choice of $n$. $c$ is a positive constant that accounts for the cost due to infection for an infected individual. 
The above cost function $\mathcal{J}(n)$ can equivalently be written as a function of $a$ as follows:
\begin{equation}\label{cost_function_a}
\mathcal{J}(a) = \min_{a \in \mathbb{R}^+,\;\; 0 \leq m(t;n) \leq1} \;\; c \int_0^T I(t;a) dt +d \frac{T}{a},
\end{equation}
 \textcolor{black}{where, $I(t;a)$ denotes the number of infected at time $t$ for a given $a$.}
The figure \eqref{cost_function_1} shows the plot of the cost function defined in relation \eqref{cost_function_a}. Figure \eqref{cost_function_1} provides valuable insights into the behavior of the cost function concerning the gap between two consecutive vaccinations. Figure \eqref{cost_function_1} shows that the cost function more or less remains at the minimum when the gap between two successive vaccinations falls within the range of 3 to 8 months. However, the plot takes an interesting turn when the gap between consecutive vaccinations exceeds 9 months. Beyond this point, the cost function begins to rise abruptly. 

\begin{figure}[ht!]
\begin{center}
		\mbox{\subfigure[]{\includegraphics[scale=0.4]{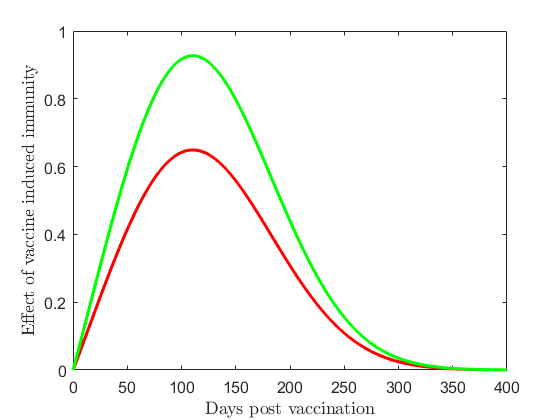}}
  \subfigure[]{\includegraphics[scale=0.4]{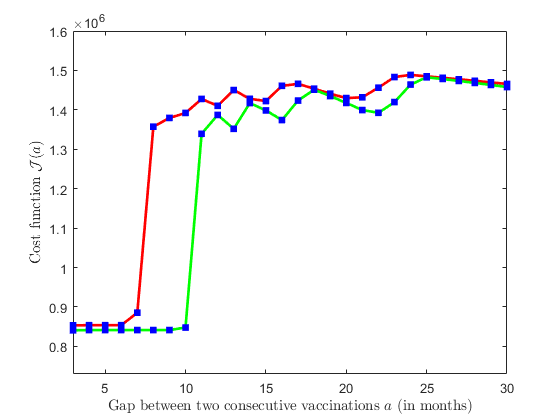}}
			}
\caption{Plot of cost function $\mathcal{J}(a)$ for different vaccine efficacy functions. Green: corresponds to formula \ref{formula_phi} and Red: corresponds to formula \ref{formula_phi} multiplied by 0.7. The parameter values: c=0.01, d=5, and all other parameter values are chosen as estimated before and as in Table.~\ref{table:estimated}.}
\label{cost_function_2}
\end{center}
\end{figure}

\begin{figure}[ht!]
\begin{center}
		\mbox{\subfigure[]{\includegraphics[scale=0.4]{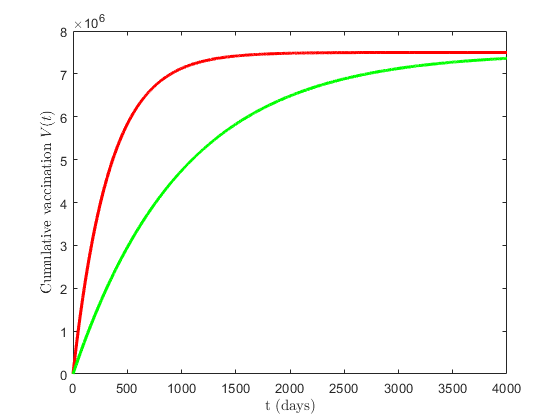}}
  \subfigure[]{\includegraphics[scale=0.4]{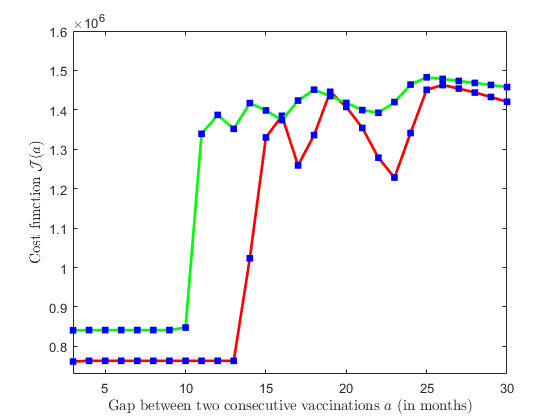}}
			}
\caption{Plot of cost function $\mathcal{J}(a)$ for different vaccination rates. The left panel corresponds to formula \ref{formula_vaccination} with the rate of vaccination $k=0.001$ (green) and $k=0.003$ (red). The right panel corresponds to the plot of the cost function with corresponding colors. The parameter values: c=0.01, d=5, and all other parameter values are chosen as estimated before and as in Table.~\ref{table:estimated}.}
\label{cost_function_3}
\end{center}
\end{figure}

\begin{figure}[ht!]
\begin{center}
	\includegraphics[scale=0.5]{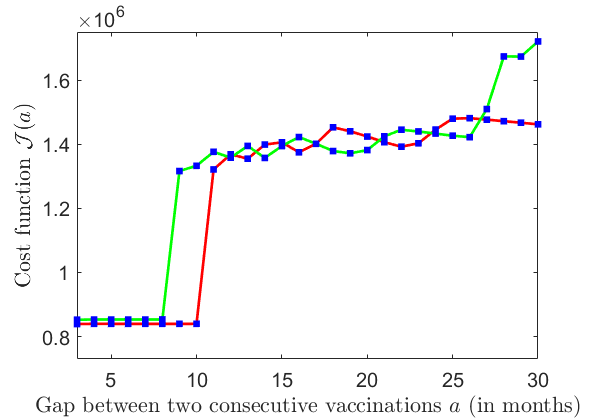}

\caption{Plot of cost function $\mathcal{J}(a)$ for different values of $\kappa$. The green and red curves correspond to $\kappa=2$ and $\kappa=1$ respectively. The parameter values: c=0.01, d=5, and all other parameter values are chosen as estimated before and as in Table.~\ref{table:estimated}.}
\label{cost_function_kappa}
\end{center}
\end{figure}

This critical observation suggests that excessively long intervals between vaccinations can be counterproductive, potentially leading to a surge in disease transmission and associated costs. This result shows that frequent vaccinations may not always be necessary and could potentially lead to diminishing returns, a phenomenon often referred to as hysteresis, whereas, an admissible larger gap between two consecutive vaccinations can effectively control the epidemic along with the minimal cost of vaccination campaign. This finding explains the significance of carefully determining the appropriate gap between two consecutive vaccination campaigns for effective epidemic control while minimizing economic burdens on a country or province.

\subsection{\bf Effect of vaccine efficacy and vaccination rate} 
A vaccination campaign focuses on two major aspects, the effectiveness of the vaccines and the rate of vaccination. These two factors can depend on the decision-makers. Thus it is important to understand the effect of vaccine efficacy and the rate of vaccination on the cost function. In figure \eqref{cost_function_2}, we plot the cost function $\mathcal{J}$ for two different vaccine efficacies. We notice that if the vaccine efficacy is less (red curve in Fig \eqref{cost_function_2}) then the cost function remain at the minimum if the gap between the successive vaccination varies between 3 to 6 months. In contrast, if the vaccine efficacy is larger (blue curve in Fig \eqref{cost_function_2}) then the cost function stays at a minimum if the gap between the successive vaccinations varies between 3 to 9 months. This observation suggests that highly effective vaccines may allow for more extended gaps between vaccinations, potentially reducing the frequency and cost of vaccination while still achieving effective epidemic control. Fig \eqref{cost_function_3} shows the plot of the cost function $\mathcal{J}$ for two different vaccination rates. This figure shows that a higher vaccination rate provides more flexibility in increasing the gap between campaigns while still controlling the epidemic effectively and minimizing costs. The results are reminiscent of the recent experiences concerning COVID-19 vaccines \cite{chattopadhyay2021infection}. This insight suggests that decision-makers should carefully consider both vaccine efficacy and vaccination rate when designing vaccination strategies to achieve cost-effective epidemic control.

\subsection{\bf Effect of co-existing strains} 
The parameter $\kappa$ accounts for the existence of multiple strains. A higher value of $\kappa$ implies the co-existence of prominent strains with very different transmission rates. Figure \eqref{cost_function_kappa} shows that a higher value of $\kappa$ provides less flexibility in increasing the gap between campaigns.

\section{Discussion and Conclusion}

\textcolor{black}{To eradicate an infectious disease through immunization, a single dose of vaccination may not be sufficient; rather, supplemental vaccine doses are essential to keep the level of immunity in the population sufficiently high over time and reduce the number of susceptible in order to achieve disease control or elimination goals \cite{WHO}. In this regard, the optimal scheduling of successive vaccination is very important, keeping in mind the cost of vaccination. In the context of diseases like Measles and Rubella, the Measles \& Rubella Initiative \cite{SIA_measles} has provided support to measles-burdened countries, focusing on sustaining high immunization coverage of children and supplementing it with supplemental doses. The comprehensive study in this work can be helpful in designing the optimal timing of successive vaccination campaigns.}

In this work, we present a comprehensive immuno-epidemic model that integrates comorbidity and the administration principle of multiple vaccine doses. The study employs a system of integro-differential equations to capture the evolving dynamics of infection and immunity over time. Considering all model parameters distributed over time-since-infection, we analyze for the dynamic changes in population immunity, determined by acquired immunity from recovered individuals and by the vaccine-induced immunity developed through multiple vaccine doses at specified intervals. The introduction of a comorbid compartment with higher infectivity allows us to implicitly capture the impact of multiple strains on the frequency and magnitude of epidemic peaks within specific time intervals.

We estimate the relevant time-distributed parameters with the help of available clinical and experimental data. Notably, our modeling results point to the substantial influence of coexisting multiple strains on both the frequency and height of epidemic outbreaks. Furthermore, we demonstrate that frequent vaccine administration may not be required and can potentially lead to a hysteresis effect on immunity levels, thus neutralizing the impact of vaccines in the longer run, sort of an anti-microbial effect. This finding challenges the conventional wisdom regarding the necessity of high-frequency vaccination strategies and indicates its negative consequence over a sustained period of administration.

\textcolor{black}{It is important to mention that, the Table 1 enlists parameter values ($N,\:V_0,\:L,\:k,\:c,\:\alpha,\:\kappa,\:b, \:\epsilon$) that are extracted from data modeling together with functions ($\phi(t),\:\psi(t),\:P(\tau),\:r(t),\:d(t),\:V(t)$) that have been implemented in contemporary references (all cited in the text, with the exception of $V(t)$). There are no known references to confirm the parameter values predicted. However, as the confidence intervals and supporting sensitivity analysis demonstrate, the model is sufficiently generic and robust against changes in parameter values, a prediction that awaits validation from future experiments.}

A critical insight arising from our findings is the paramount importance of determining the optimal gap between two consecutive vaccine doses. We emphasize that this determination should be driven by a meticulous analysis of the evolving level of immunity within the population. To address this, we propose an optimal control strategy aimed at minimizing both the number of infections and the economic costs associated with vaccination efforts. This approach underscores the need for a tailored and strategic vaccination plan that considers the interplay of factors such as population immunity, vaccine efficacy against different strains, and the cost-effectiveness of each vaccination effort. In summary, our primary objective has been to contribute to the understanding of the optimal gap between two consecutive vaccinations. Our findings highlight the complexity of disease control strategies, advocating for a more nuanced and adaptable approach that accounts for the interplay of immunity dynamics, multiple strains, and vaccination frequency in shaping effective public health interventions.

The model presented in this study is generic and applicable to various epidemic diseases. In this specific analysis, we have resorted to simplifying assumptions like homogeneous propagation, absence of ethnic migration, and patient compartmentalization, etc. that can be subjectively assessed against specific data through minor modifications of this model. Our assumption of equal effectiveness across all vaccine doses and the lack of differentiation between individuals who received the first, second, or booster dose are the limitations of the present work. Future research could enhance the model's precision by incorporating more accurate data on vaccine efficacy and distinguishing between different doses to calculate immunity levels more effectively. Additionally, the model's assumption of a homogeneous population, with individuals sharing similar immune statuses, overlooks potential variations influenced by factors like age. Subsequent investigations could enrich the model by introducing population heterogeneity through age-structured modeling, providing a more realistic representation of the epidemic dynamics.



\section{Conflicts of interest}

The authors declare that they have no conflict of interest.

\section{Data availability}

All data generated and analyzed during this study will be curated in the Aston University repository and made available on request.

\bibliographystyle{unsrt}
\bibliography{accepted}

\end{document}